\def\plotfiddle#1#2#3#4#5#6#7{\centering \leavevmode
\vbox to#2{\rule{0pt}{#2}}
\includegraphics{#1}}
\documentstyle{mn}
\title[Line Width Distributions in AGN]{Line width distributions as evidence for axisymmetry in the broad line regions of active galaxies} 
\author[C.M.Rudge \& D.J.Raine]{C.M.~Rudge$$ and D.J.~Raine$$
\\$$Astronomy Group, University of Leicester, University Road,
Leicester, LE1 7RH, UK.}
\begin{document}
\maketitle
\begin{abstract}
The nuclei of a wide class of active galaxies emit broad emission
lines with widths at half maximum (FWHM) in the range $10^{3}-10^{4}$ km
s$^{-1}$.  This spread of widths is not solely a consequence of the
range of the luminosities of these sources since a plot of width
versus luminosity shows a large scatter. We propose that the broad
line emission region (BLR) is axially symmetric and that this scatter
in line width arises from an additional dependence on the angle of the
line of sight to the axis of the emission region.  Such a relation is
natural in unified models of active nuclei which link a variety of
observed properties to viewing angle. Adopting a simple form for the
line width as a function of luminosity and angle, and convolving this
with the observed luminosity function, allows us to predict a line
width distribution consistent with the available data. Furthermore, we
use the relation between the equivalent width of a line and the
luminosity in the continuum (the `Baldwin Effect') to predict an
observed correlation between line width and equivalent width. The
scatter on this correlation is again provided by angular
dependence. The results have applications as diagnostics of models of
the broad line emission region and in cosmology.
\end{abstract}
\section{Introduction}

In unified models of active galactic nuclei with spherically symmetric
BLR the width distribution of the broad emission lines cannot be
accounted for by luminosity dependence alone. Plots of line width
versus continuum luminosity have a large scatter and show no
significant correlation \cite{W93,P97}. There is, however, growing
evidence that the broad line region (BLR) is not spherical, but
axisymmetric.
\begin{enumerate}
\item Observed samples of AGN
\cite{Wills86,W93,Brotherton94} suggest relations between
line widths and R, the ratio of core to lobe dominance.
Other samples \cite{P97} find relations between line widths
and $\alpha_{\rm ox}$, the continuum slope parameter from
the optical to X--ray bands. Both of these parameters have some
viewing angle dependence.
\item The continuum and line light curves of some active
nuclei, eg 3C390.3 \cite{Wamsteker97}, are most naturally
interpreted in terms of a disc-like line emission region.
It has also been suggested that some double peaked line
profiles arise from discs (eg Arp 102b) although the
interpretation in these cases is not so clear when time
variability is taken into account.
\item Several axisymmetric disc-wind models, such as those
of Cassidy \& Raine \shortcite{Cassidy96}, Chaing \& Murray
\shortcite{Chaing96} and Emmering, Blandford \& Shlosman
\shortcite{Emmering92} have been proposed and  models of
this type are gaining support from evidence for winds
\cite{Pasadena}.  These will naturally predict some viewing
angle dependence of line width.
\end{enumerate}
It should be noted that Osterbrock \shortcite{Osterbrock77} showed
that a deficit of systems with narrower lines ruled out pure disc
models, but such objections do not necessarily apply to axisymmetric
models in general. 

In this paper we shall adopt a simple dependence of line width on both
viewing angle and luminosity. Then: 
\begin{enumerate}
\item We obtain a reasonable fit to the distribution of line widths.
\item Given the Balwin relation between line and continuum
luminosity we predict a relation between line width and
equivalent width compatible with the observed trend. The
scatter on this relation is attributed to angular
dependence.
\item We discuss how the width distribution can be used to
test models of the BLR.
\item If the BLR is indeed axisymmetric, we show how the
line width distribution can be used, in principle, to
determine cosmological parameters.
\end{enumerate}

\section{Modelling the Line Width Distribution}
For an axisymmetric BLR we can expand the line width as a function of
angle $i$ between the axis and the line of sight in a Fourier series
\[v(i) = a_{0}+\sum_{n=1}^{\infty} a_{n}\sin ni \]
with the cosine terms set to zero for symmetry about
$i=\frac{\pi}{2}$.  However the line widths do not appear to be a
function of angle only as it is difficult to obtain the range of
widths required in an uncontrived way.  Also, observations of variable
sources such as NGC 4151 \cite{Ulrich91}  show that line width is
luminosity dependent. However, there is no reason to assume that the
variation in line width in one source should be the same as that
between sources at different luminosities. In fact in the case of line
equivalent widths this is definitely not the case
\cite{Peterson}. Therefore we introduce the luminosity dependence as
another independent variable.  Thus, we consider $v(i,L)$ to depend on
angle and  luminosity with $a_{n}=a_{n}(L)$, which we expand as
\[a_{n}(L) = L^{\alpha}\sum_{m=0}^{\infty} a_{nm}L^{m}\ {\rm for}\ n\geq0.\]
Note that in principle $\alpha$ could depend on $n$, but we
shall take it to be a constant for a given line.  To lowest
non-trivial order we terminate the  series at $a_{10}$ and write
$a_{00}=a$, $a_{10}=b.$  Thus, we arrive at our proposed model:
\[v(i,L)=(a+b\sin i)L^{\alpha}. \] 
More precisely, we suggest that line width at half maximum (FWHM), $v$
km s$^{-1}$, is related to the B band luminosity, in units of
$10^{44}$ erg s$^{-1}$, and viewing angle as follows
\begin{equation}
v(i,L_{44})=(a+b\sin i)L_{44}^{\alpha}.
\label{vileqtn}
\end{equation}

To produce the line width distribution we combine the above relation
with statistical information about the number of systems at each
viewing angle and each luminosity. The number of systems, $T$, at each
viewing angle satisfies 
\[{\rm d} T(i)=\sin i \ {\rm d}i,\  0 < i < i_{\ast} < 90^{\circ} .\]  
The angle $i_{\ast}$ takes account of the obscuring material in the
unified picture that prevents direct sight of the BLR at large
inclinations. This value will vary between systems and may be
correlated with luminosity. Realistic unified models probably have
$i_{\ast}=i_{\ast}(L_{44})$ but we shall take $i_{\ast}=$ constant,
since this is all that can be justified by current data.

The number of systems at each luminosity is found from the luminosity
function $\Phi(L)$. We use a broken power law fit to the luminosity
function given in Boyle, Shanks and Peterson \shortcite{Boyle88} to
give \[\Phi(L_{44}) \propto L_{44}^{-p} \hspace{6mm}
\left\{\begin{array}{ll} p=3.85\ \rm{for} \ L_{44} > 2.0\ \\
p=1.27\ \rm{for}\ L_{44}<2.0 \end{array} \right. .\]  The number of
systems at each width, $N(v)$, is given by
\[ N(v) = \int_{0}^{\infty}\int_{0}^{i_{\ast}} \delta(v-v(i,L_{44})){\rm d}T(i) \Phi(L_{44}){\rm d}L_{44} \]
wher the $\delta$ -- function restricts the integral over luminosity
and angle to systems with width $v$. Thus evaluating the angular
integral, we obtain
\begin{equation}
N(v) = \int_{0}^{\infty} \frac{\sin i}{\left | \frac{{\rm d}v}{{\rm d}i} \right |} \Phi(L_{44}) {\rm d}L_{44}.
\end{equation}
To compute this integral numerically we restrict $L_{44}$ to the range $0.01 < L_{44} < 13.0$.

\subsection{Equivalent widths}
To complete the picture we introduce the parameter $\beta$ for the
Baldwin  relation between the equivalent width (EW) of a line and the
blue band continuum luminosity of the active nucleus:
\[ EW \propto L_{44}^{\beta} .\]
From the observational data sets we can obtain a relation between the
EW and FWHM of a line of the form \[ EW\propto v^{\gamma} .\] Using our
previous relation, $v \propto L_{44}^{\alpha}$, we see 
\begin{equation}
EW \propto v^{\beta/\alpha}.
\end{equation}
with a scatter deriving from the angular dependence.  Thus, any two of
$\alpha$, $\beta$ or $\gamma$ should determine the third.
In particular, given the Baldwin relation and a fit to $N(v)$, determining
$\alpha$, we predict the slope of the correlations between FWHM and EW (with a
scatter from angular dependence).

\section{Results}\subsection{Line width distribution}
\begin{figure*}
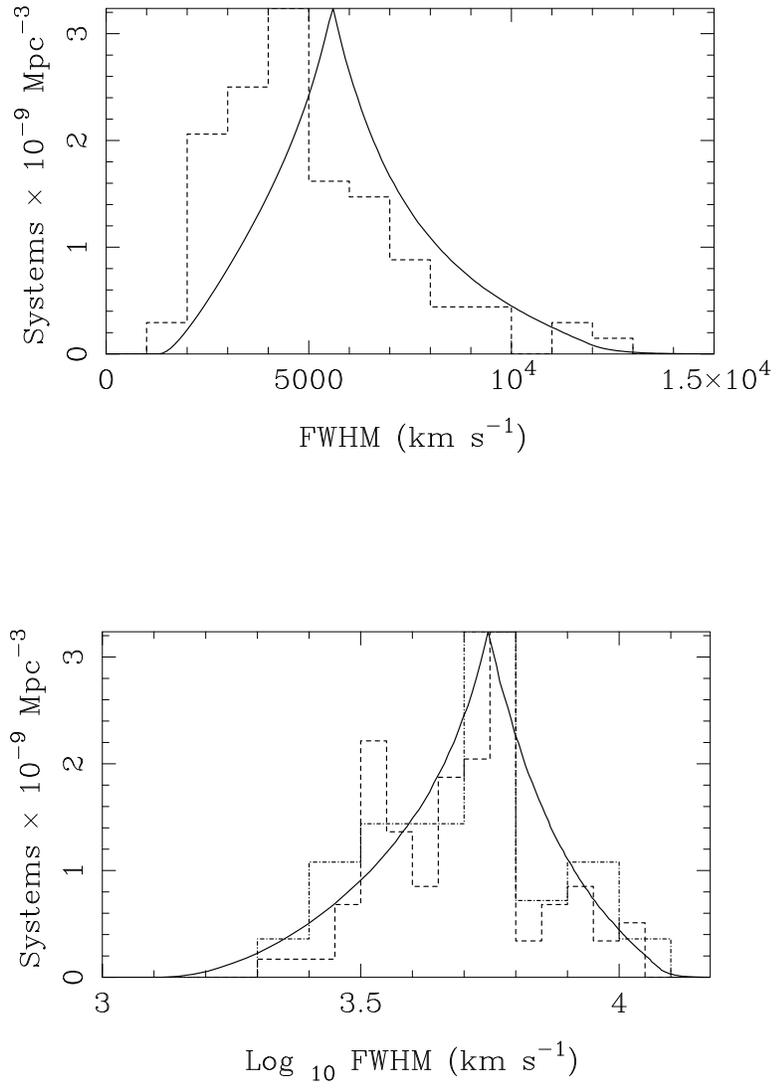

\plotfiddle{c4wilkes.eps}{225pt}{-90}{40}{40}{-150}{225}
\plotfiddle{c4wandb.eps}{225pt}{-90}{40}{40}{-150}{225}
\caption{Model predictions for C\,IV overlaid
with observed data by Wilkes (1987) (top) and
Wills et al.\ (1993) (dashed)
and Brotherton et al.\ (1994) (dot-dashed).}
\label{c4lwd}
\end{figure*}

\begin{figure*}
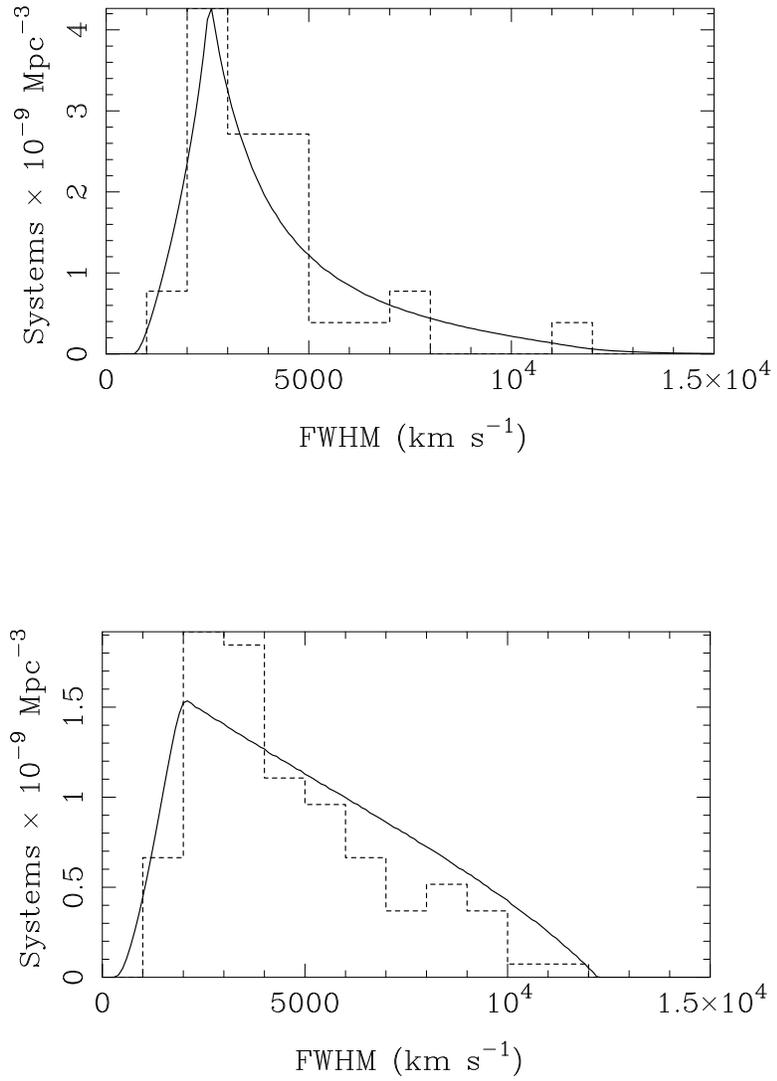

\plotfiddle{rixoshb.eps}{225pt}{-90}{40}{40}{-150}{225}
\plotfiddle{rixosmgii.eps}{225pt}{-90}{40}{40}{-150}{225}
\caption{Model predictions for H$\beta$ (top) and
Mg\,II (bottom) overlaid with observed data
from Puchnarewicz et al.\ (1997).}
\label{rixosfits}
\end{figure*}

We have investigated the C\,{\scriptsize IV} $\lambda$1549,
Mg\,{\scriptsize II} $\lambda$2798 and H$\beta\ \lambda4861$ lines
using data from Wills et al.\ \shortcite{W93}, Brotherton et al.\
\shortcite{Brotherton94}, Wilkes \shortcite{Wilkes87} and Puchnarewicz
et al.\ \shortcite{P97}.  Fig.\ref{c4lwd}\ shows the observed width
distribution in C\,{\scriptsize IV} for three samples with a
theoretical curve drawn for comparison using $i_{\ast} = 60^{\circ} $
and the other parameters as shown in table 1. This value for
$i_{\ast}$ is chosen to give a reasonable fit for the three emission
lines simultaneously. This value is larger than expected for the
standard unified model of AGN but consistent with the unified model of
Cassidy and Raine \shortcite{Cassidy97}. The position of the peak of the
distribution is dependent mostly on $a, b$ and $i_{\ast}$ with
$\alpha$ determining the spread of the distribution. Setting $a=0$
gives a pure disc model and increases the predicted number of narrower
lined systems. Setting $b=0$ gives a spherically symmetric BLR model
and the distribution is a narrow peak centred around $a$ (assuming
$\alpha$ is small). The Wilkes data, taken from the Parkes flat
spectrum survey, is biased towards face on objects, which this model
predicts would give a peak in the distribution at lower velocity.  The
data in Wills et al.\ and Brotherton et al.\ is biased towards higher
luminosity objects which may explain the lack of narrower lined
systems in the lower figure.  Fig.\ref{rixosfits} shows similar fits
to Mg\,{\scriptsize II} and H$\beta$.

\begin{table}
\caption{Model parameters used to fit the observed samples for 3
emission lines. Value of $\beta$ taken from Peterson (1997) for C\,IV and
from Puchnarewicz, E.M. (private communication) for Mg\,II and H$\beta$}
\begin{tabular}{c r r r r}
Line &
\multicolumn{1}{c}{a} & 
\multicolumn{1}{c}{b} & 
\multicolumn{1}{c}{$\alpha$} &
\multicolumn{1}{c}{$\beta$} \\
C\,{\scriptsize IV} & 2500 & 9500 & 0.15 & -0.17 \\
H$\beta$ & 2500 & 8000 & 0.30 & 0.4 \\
Mg\,{\scriptsize II} & 500 & 2500 & -0.35 & -0.37 \\
\end{tabular}
\label{parameters}
\end{table}

\subsection{FWHM - EW}
\begin{figure*}
\plotfiddle{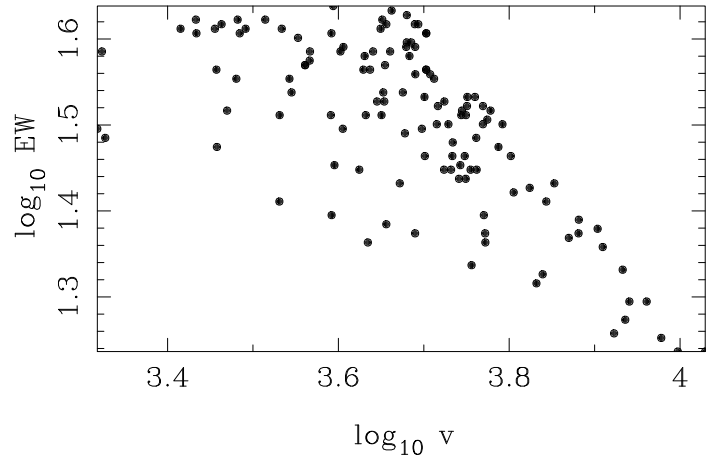}{250pt}{-90}{40}{40}{-150}{250}
\caption{FWHM - EW relation as predicted by this model for
C\,IV. Notice the large scatter produced by the
angular dependence of $v$. The sharp edge is expected to be
disrupted by source variability.}
\label{WEWplots}
\end{figure*}

Fig.\ref{WEWplots} shows the corresponding EW -- FWHM relations
obtained from the model. The slopes, from table \ref{parameters}, are
$\beta/\alpha = -0.85, 1.1, 0.89$  for C\,{\scriptsize IV},
Mg\,{\scriptsize II} and H$\beta$ respectively compared with
$-1\pm$0.25 for C\,{\scriptsize IV} (from inspection of Wills et al.\
\shortcite{W93}), 1.6$\pm$0.4 for Mg\,{\scriptsize II} and 0.6$\pm$
0.4 for H$\beta$ (Puchnarewicz, E.M., private communication). The
sharp edge to the data in these figures will be disrupted by time
variability in the source luminosity since the Baldwin relation in a
single variable source does not have the same slope as the relation
between different sources \cite{Peterson}.

\section{Discussion}
\subsection{Mg\,II an exception?}
Whilst it is perhaps unexpected for the Mg\,{\scriptsize II} emission
line distribution to be fitted with $\alpha < 0$, this is necessary
for consistency with the observed Baldwin effect ($\beta <0$) amd the
positive correlation between EW and FWHM ($\beta/\alpha >0$). Although
a positive $\alpha$ gives a similar distribution curve, $\alpha<0$ is
physically more reasonable as it corresponds to a decrease in Mg$^{+}$
ions with increasing ionization parameter. Systems which have
comparably wide Mg\,{\scriptsize II} and C\,{\scriptsize IV} lines are
rare and seem to occur at low luminosity (e.g. NGC 4151).  More
commonly the widths $v_{\rm{line}}$ satisfy $v_{{\rm Mg\,{\scriptsize
II}}} \ll v_{{\rm C\,{\scriptsize IV}}}$. Presumably at high
luminosity the low ionization emission lines are absent from the
inner, faster moving, BLR material. Certainly at higher luminosity
(e.g. NGC 5548) reverberation mapping shows Mg\,{\scriptsize II} to
come predominantly from larger radii \cite{Clavel91}.

\subsection{Narrow lines}
Narrow lined systems (FWHM$<500$ km s$^{-1}$) appear to be
under-represented in the samples. This may be a resolution or a
selection effect. However a pure disc model ($a=0$) would predict
larger numbers of such systems. This led Osterbrock
\shortcite{Osterbrock77} to rule out disc models for the BLR. However, it is
as yet unclear where the  Narrow Line Seyfert 1 class of objects  fits
into the scheme. Recent analysis of these systems \cite{R-P97} shows
the existence of a broad component to the narrow lines suggesting that
the BLR is not obscured in these sources, which may therefore provide
a class of ``broad lined'' objects with small FWHM.

\subsection{Comparison with theoretical axisymmetric models}
\begin{figure*}
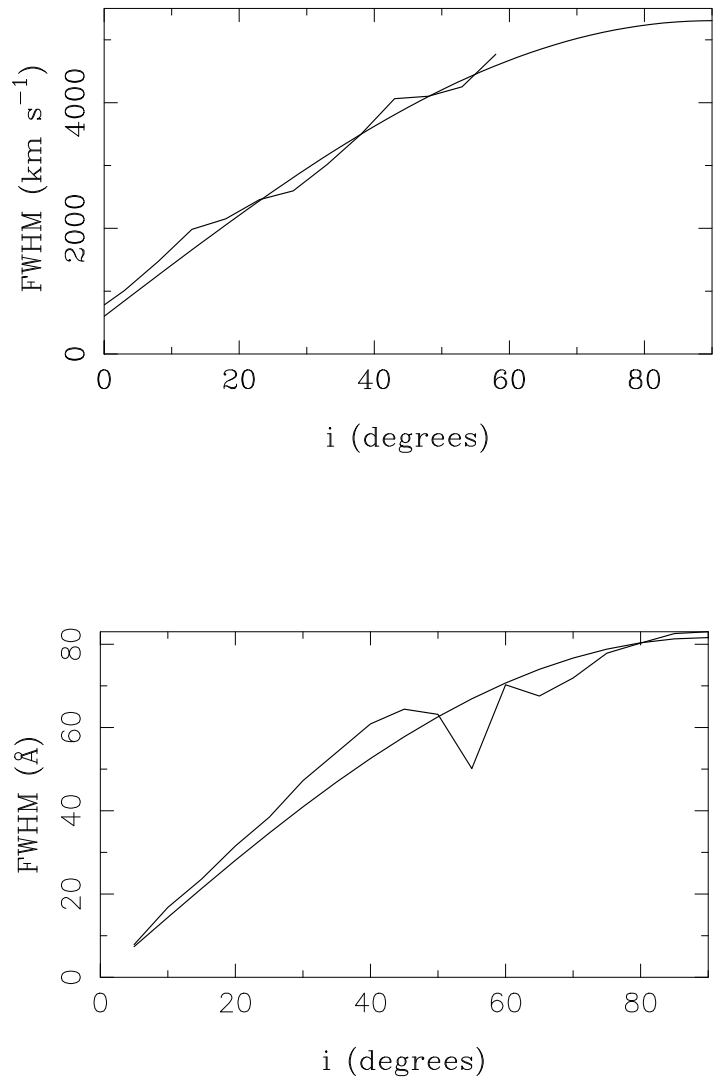

\plotfiddle{crsincurve.eps}{225pt}{-90}{40}{40}{-150}{225}
\plotfiddle{cmsincurve.eps}{225pt}{-90}{40}{40}{-150}{225}
\caption{FWHM - viewing angle relation predicted by the dual
winds model (top) and the disc wind model (bottom). Both
models are fitted well by a sine
curve.}
\label{sincurves}
\end{figure*}

Current axisymmetric models of the BLR include the dual winds picture
\cite{Cassidy96}, the disc wind model \cite{Chaing96} and the magnetic
wind model \cite{Emmering92}. The dual winds model and the disc wind
model both have an angular dependence of FWHM on viewing angle which
is consistent with our model (Fig.\ref{sincurves}). Examination of
figure 6 in Emmering et al.\ \shortcite{Emmering92} suggests that in
the magnetic wind model the FWHM reduces as viewing angle
increases. Whilst this is in the opposite sense to the other models we
can still obtain the line width distribution using equation
(\ref{vileqtn}) with $b<0$ and so cannot rule out the magnetic winds
model. However where correlations exist between FWHM and R or
$\alpha_{ox}$, they suggest that $v$ increases with $i$, hence
$b>0$. It may be possible to discriminate between the models by
consideration of the predicted value of $\alpha$.

\subsection{Application to cosmology}
Since this analysis enables us to remove statistically the scatter in
the luminosity--FWHM relation for AGN it has obvious
applications to cosmology once large data sets of FWHM become
available. Note that the observed distribution of line widths is
itself independent of any cosmological model. If other methods can be
used to determine the line of sight to individual objects these too
could then be used for cosmology. In further work we expect to include
a redshift dependence in the luminosity function $\Phi(L,z)$ \cite{Pei}
with $L=L(H_{0},q_{0})$, to provide a new method of determining $H_{0}$
and $q_{0}$.

\section{Conclusions}
We conclude that the simple picture we have presented here accounts
for the  scatter in FWHM versus luminosity, accounts for the
distribution of FWHM, and relates the trend in the EW with FWHM to the
Baldwin relation. This may be useful as a diagnostic tool in discriminating between disc-wind models. The analysis has
applications as a cosmological tool particularly as measurement of
line widths is independent of any cosmological model.

\section{acknowledgements}
The authors wish to thank Sarah Symons and Richard Thorpe for their
contributions at an early stage of this work; also E.M. Puchnarewicz
for providing data and helpful comments. CMR acknowledges the support
of PPARC, in the form of a research studentship.
\newpage

\end{document}